\documentstyle[12pt]{article}
\input{psfig.tex}
\begin{document}
{\rightline{IMSc-96/09/28}}
\begin{center} 
{\large{\bf On a Static solution of Einstein Equations with incoming and outgoing
radiation }} 
\end{center} 
\vskip 1.0cm
\begin{center}
{\large G. Date\footnote{ Institute of Mathematical Sciences, CIT Campus, Tharamani,
 Madras - 600 113, India.  \par e-mail:~shyam@imsc.ernet.in}
}
\end{center}
\vskip 1.0cm
\begin{abstract}
Einstein equations with $T_{\mu\nu} = k_\mu k_\nu + \ell_\mu \ell_\nu$
where $k, \ell$ are null are considered with spherical symmetry and
staticity. The solution has naked singularity and is not asymptotically
flat. However, it may be interpreted as an envelope for any static
spherical body making it more massive. Such an interpretation and some of its 
implications are detailed.
\end{abstract}
\newpage
\centerline{ {\bf Introduction }}
\vskip .5cm
Any astrophysical body has a temperature and thus is a source of
outgoing radiation. Equally well, every such body also receives an
incoming radiation eg. the cosmic background radiation. Depending upon
the respective temperatures there will either be a net outgoing or
incoming flux of radiation. It is then conceivable that the two rates
exactly match and one reaches an equilibrium situation. In such a
situation, the stress tensor in the vicinity of such a body may be taken
to be of the form 
\begin{equation}
T_{\mu\nu} \ = \ k_\mu k_\nu + \ell_\mu \ell_\nu \ , \ k^2 = \ell^2 = 0 \ , \ k. \ell > 0 
\end{equation}
\par
When the rates are unequal either the $k$ or the $\ell$ term may be taken
to be dominating and one essentially gets the Vaidya solution (non stationary)
\cite{vaidya}, or collapsing null fluid shell case.
However when the rates are precisely matched, both terms are important
and one can look for a static solution. As a first step in this
direction of course one can consider the simpler case of spherically
symmetric solution.

In the present work we consider such a solution. Some of the salient
features of the solution are the following.

Let $r$ denote the usual Schwarzschild radial coordinate.

\begin{enumerate}
\item As $r \rightarrow 0$ the solution has a curvature singularity
which is naked (i.e. no event horizon).
\item As $r \rightarrow \infty$, the metric components go as $\ln(r)$
and thus the solution is {\it not} asymptotically flat.
\item These two features make it difficult to interpret the solution
physically. However, one can consider the solution to be valid for $R
\le r \le \bar{R}$ range. At $R$ one can match the solution for a typical
interior Schwarzschild solution while at $\bar{R}$ one can match it with the
standard exterior Schwarzschild space-time.
The asymptotics of the solution permit such a
$(C^0)$ matching. If $M$ denotes the mass of the interior solution and
$\bar{M}$ denote the mass indicated by the matching at $\bar{R}$, thus $\bar{M} > M$
and therefore the mass measured (deduced) from $r >> \bar{R}$ is larger than
$M$. One can consider the matching at $R$ just outside a black hole or even 
with a negative mass Schwarzschild solution, but $\bar{M}$ is always positive
for sufficiently large $\bar{R}$.
\end{enumerate}

The paper is organised as follows.

Section 1 contains basic equations which are straight forward to derive
In Section 2, we present analysis of qualitative features of the
solution. The asymptotics are discussed and numerical solutions are
presented corroborating the qualitative analysis. In Section 3 we discuss
some of the possible matchings and summarise our conclusions. The
appendix contains a few details of general spherically symmetric and static
non-empty space time. A some what similar {\it{ exact solution}} is also 
presented and the details of matchings are discussed.\\
\vskip .5cm
\centerline{ \bf{Section 1: Basic equations} }
\vskip .5cm

The basic equations are (Signature + - - - )
\begin{equation}
R_{\mu\nu} - \frac{1}{2}  Rg_{\mu\nu} \ = \ 8\pi T_{\mu\nu}
\end{equation}

\begin{equation}
T_{\mu\nu} \equiv \rho k_\mu k_\nu + \sigma \ell_\mu \ell_\nu \ , \ k^2
\ = \ \ell^2 \ = \ 0 \ , \ k.\ell > 0\,.
\end{equation}

where $k^\mu, \ell^\mu$ vector fields represent massless radiation
outgoing and incoming respectively. 

Clearly $g^{\mu\nu} T_{\mu\nu} =0$ and therefore $R$ term can be
dropped. Using spherical symmetry and staticity we write 

\begin{equation}
ds^2 \ = \ F(r) dt^2 - G(r) dr^2 - r^2 (d\theta^2 + \sin^2 \theta
d\phi^2)
\end{equation}

The killing vectors are : \\

\begin{equation}
\begin{array}{rcl} 
\frac{\partial}{\partial t} & \rightarrow & \mbox{generating time translations ;} \\
& & \\
\xi(\alpha, \beta, \gamma) & \equiv & \xi^\theta \partial_\theta + \xi^\phi
\partial_\phi ~~~~\mbox{where,} \\
& & \\
\xi^\theta & = & \alpha \sin \phi \ + \ \beta \cos \phi \\
\xi^\phi & = & \gamma + cot \ \theta (\alpha \cos t \phi - \beta \sin \phi) ,\\
& & \\
\end{array}
\end{equation}

generating isometries implied by spherical symmetry.

It follows immediately that
\begin{equation}
{\cal {L}}_\xi T_{\mu\nu} \ = \ 0 \quad \leftrightarrow \quad {\cal {L}}_\xi k_\mu \ = \
{\cal {L}}_\xi \ell_\mu \ = \ 0
\end{equation}

Combining with $k^2 \ = \ \ell^2 \ = \ 0$ one gets
\begin{equation}
\begin{array}{rclcrcl}
k^\mu & = & k^0 (1, \sqrt{\frac{F}{G}}, 0,0) &, & \ell^\mu & = & \ell^0 (1, -
\sqrt{\frac{F}{G}}, 0, 0) \\
k^0, ~~ \ell^0 & > & 0 & , & 2k^0 \ell^0 F & > & 0
\end{array}
\end{equation}

Here $\mu = 0,1,2,3 \leftrightarrow t,r,\theta,\phi$ respectively. Since
$k^0, \ell^0$ are arbitrary at this stage, we absorb $\rho$ and $\sigma$ in
$k^0 ~,~ \ell^0$ respectively.

We can actually solve for $k^0$ and $\ell^0$ using the conservation
equations. 

\begin{equation}
\begin{array}{ccl}
\nabla_\nu \ T^{\mu\nu} & = & 0 \quad \Rightarrow \\
(k^\mu \nabla \cdot k + k \cdot \nabla k^\mu) \ + \ (\ell^\mu
\nabla \cdot \ell + \ell \cdot \nabla \ell^\mu) & = & 0\,. \\
\end{array}
\end{equation}

Evaluating the covariant derivatives etc. shows that each bracket is
separately zero implying that $k^\mu$ and $\ell^\mu$ integral curves are
geodesics (non affinely parametrised). Furthermore, one gets
\begin{equation}
k^0 \ = \ \frac{B_+}{rF} \ , \ \ell^0 \ = \ \frac{B_-}{rF} ,~~~~{\mbox{ where
$B_\pm$ are constants}}.
\end{equation}
Therefore,
\begin{equation}
k^{\mu} \ = \ \frac{B_+}{rF} \ \left( 1, \sqrt{\frac{F}{G}} , 0, 0\right) 
\ell^\mu \ = \ \frac{B_-}{rF} \ \left( 1, -\sqrt{\frac{F}{G}} , 0, 0\right)
\end{equation}

The nonvanishing components of $T_{\mu\nu}$ then are
\begin{equation}
T_{00} \ = \ \frac{B^2_+ + B^2_-}{r^2} \ , \ T_{11} \ = \ \frac{G}{F}
T_{00} \ , \ T_{01} \ = \ \sqrt{\frac{G}{F}} \left( \frac{B^2_- -
B^2_+}{r^2}\right)
\end{equation}

For spherically symmetric, static metric $R_{01}$ is zero and therefore
$B^2_+ = B^2_-$ and  since $k^0, \ell^0$ are both positive we take $B_+
= B_- \equiv B~$. Then,

\begin{equation}
T_{00} \ = \ \frac{A}{r^2} \ , \ T_{11} \ = \ \frac{A}{r^2} \
\frac{G}{F}
\end{equation}

\begin{equation}
k^0 \ = \ \ell^0 \ = \ \frac{B}{rF} \ , \ A \ \equiv \ 2B^2 
\end{equation}

Note that in the geometrised units we are using $T$ has dimensions of
(length)$^{-2}$ and thus $A$ is dimensionless. Since $T_{00} \sim
\frac{1}{r^2}$ we see both that we cannot have asymptotic flatness and
that there will be a curvature singularity at $r=0$.

It is straightforward to compute $R_{\mu\nu}$. Out of the 10 Einstein
eqns. the eqns for $(\mu\nu) = 02,03,12,13,23$ are identically satisfied.
The 01 equation has already been used to set $B_+ = B_-$. The 22 and 33
equations are identical. This leaves us with 3 nontrivial equations.
Setting $f \ \equiv \sqrt(F) \, \ g \ \equiv \sqrt(G) \ $, the equations
are:
\begin{equation}
{\mbox{
$\begin{array}{lccl}
(00) : & g^{-1} \left( \frac{f'}{g}\right) ' \ + \ 2r^{-1} \frac{f'}{g^2} 
& = & 8\pi f^{-1} \frac{A}{r^2}  \\
(11) : & -f^{-1} \left( \frac{f'}{g}\right) ' \ + \ 2r^{-1} \frac{g'}{g^2} 
& = & 8\pi gf^{-2} \frac{A}{r^2} \\ 
(33) : & -f^{-1} \frac{f'}{g^2} \ + \ g^{-1} \frac{g'}{g^2} \ - \ r^{-1} \left( \frac{1}{g^2}-1 \right) 
& = & 0 \\
\end{array}$
}}
\end{equation}

\vskip .5cm

\noindent {\bf Remarks}

\begin{enumerate} 
\item $A=0$ gives the standard Schwarzschild case. The 33 eqn is independent of
 $A$ and is invariant under constant rescaling of $f$.
\item Under $f \rightarrow \lambda f$ the 00 and 11 equations retain their form but
with $A \rightarrow A / \lambda^2 $
\item All the equations are also invariant under $r \rightarrow \lambda r$ and
therefore there is no intrinsic scale available at the level of the equations. 
This has to be provided by physical boundary conditions. This is of course true 
for Schwarzschild case as well (indeed whenever the stress tensor is traceless
and has no dimensionful parameters).
\item By suitable combinations at 00 and 11 equations we get one equation which contains $A$
dependent term while another one which does not contain explicit $A$ dependence. The
$A$ independent combination is a second order differential equation while $A$
dependent one is a first order equation. The 33 equation is 1st order.
\end{enumerate}

It is straightforward to verify that the second order equation is automatically
satisfied if the two first order equations are satisfied.

Defining $r = \mu_0 e^{\xi}$, $\mu_0$ an arbitrary scale and $\lambda \equiv
8\pi A$, we write the equations in terms of $F$ and $G$, as :

\begin{equation}
\frac{dF}{d\xi} ~~ = ~~ \lambda G \ + \ F(G-1) 
\end{equation}
\begin{equation}
\frac{dG}{d\xi} ~~ = ~~ \lambda \ \frac{G^2}{F} - G(G-1) 
\end{equation}
\begin{equation}
\frac{2}{F} \frac{d^2F}{d\xi^2} \ - \ \left( \frac{1}{F} \frac{d}{d\xi} F\right)^2 -
\frac{dF}{Fd\xi}\frac{dG}{Gd\xi} \ - \ \frac{2}{G} \frac{dG}{d\xi} ~~ = ~~ 0
\end{equation}

The last equation is the second order equation which is identically satisfied 
if the first two equations 
hold. Thus the basic equations to be solved are the first two equations.

To summarise : \\

\begin{center}
$\begin{array}{lcl}
ds^2 & = & F dt^2 - Gdr^2 - r^2 d\Omega^2  \\
& & \\
k^0 & = & \frac{\sqrt{A/2}}{rF} ~~~~~~~~ (~ = ~\ell^0) \\
k^\mu & = & k^0 (1, \sqrt{\frac{F}{G}},0,0) \\
\ell^\mu & = & k^0 (1, - \sqrt{\frac{F}{G}} , 0,0) \\
& & \\
r & \equiv & \mu_0 e^{\xi}~ , ~ \prime ~ \leftrightarrow ~ \frac{d}{d\xi} \\
& & \\
F' & = & \lambda G \ + \ F(G-1) \\
G' & = & \lambda \frac{G^2}{F} - G(G-1) \\
\end{array}$
\end{center}

The case $\lambda=0$ gives the Schwarzschild solution. 

For $\lambda \ne 0$, we set $F ~\equiv~ \lambda \Phi$. The equations for $G$
and $\Phi$ then have no $\lambda$ dependence and are:
$$
\Phi' \ = \ G + \Phi(G-1) \eqno(A) 
$$
$$
G' \ = \ \frac{G^2}{\Phi} -G(G-1) \eqno(B) 
$$
In the next section we will analyse these equations.

\newpage
%
\centerline{ \bf{Section 2: Analysis of the equations}}
\vskip .5cm

\begin{equation}
{\mbox{
$\begin{array}{cclcccl}
\Phi & > & 0 & , & G & > & 0 \\
\Phi' & = & G + \Phi(G-1) & , & G' & = & \frac{G^2}{\Phi} -G(G-1) 
\end{array}$
}}
\end{equation}
Therefore,
$$
G \ = \ \frac{\Phi+\Phi'}{\Phi+1}
$$
Eliminating G from the second equation, gives a second order equation
for $\Phi$, namely,
\begin{equation}
\Phi''\{\Phi(\Phi+1)\} + \Phi'\{\Phi(\Phi-2)\} - \Phi'^2 -2\Phi^2 \ = \ 0\,. 
\end{equation}
This can be integrated once to give,
\begin{equation}
\Phi'(1+\frac{1}{\Phi}) + \Phi - 2\ln (\Phi) - 2\xi \ = \ C
\end{equation}
or,
\begin{equation}
\Phi' \ = \ (C-\Phi+2\ln (\Phi) + 2 \xi ) (\frac{\Phi}{\Phi+1})
\end{equation}
Substituting in the expression for $G$, we get the exact equations: 
\begin{equation}
{\mbox{
$\begin{array}{cclc}
G & = & \frac{\Phi}{(\Phi+1)^2} [C+1 + 2\ln (\Phi) + 2\xi] & 
{\mbox{(solves G in term of $\Phi$)}}
\\
& & & \\
\Phi' & = & \frac{\Phi}{(\Phi+1)} [C-\Phi + 2\ln (\Phi) + 2\xi] &
{\mbox{(1st order equation for $\Phi$)}}                
\end{array}$
}}
\end{equation}
The first integral has given us {\it one} constant of integrations.

\noindent {\it Remarks:} 

1.	\ $G \equiv 0, \ \Phi' = -\Phi$ is an exact solution. This
follows from both the original equations for $\Phi$ and $G$ and from the above
equations. However, this is {\it not} an acceptable solution.

2.	\ If $\Phi'=0$ then
\begin{equation}
C + 2\ln (\Phi) + 2\xi \ =\ \Phi
\end{equation}
\begin{equation}
\Phi'' \ = \ \left( \frac{\Phi}{\Phi+1}\right)' [~~~~~] + \left(\frac{\Phi}{\Phi+1}\right) \left[ -\Phi' +
\frac{2}{\Phi} \Phi' +2\right]
\end{equation}
Therefore,
\begin{equation}
\Phi'' |_{\Phi'=0} \ = \ 0 + \frac{2\Phi}{\Phi+1} \ > 0 
\end{equation}
Thus $\Phi$ has at the {\it {most one minimum}}.
The minimum is determined by
\begin{equation}
\Phi(\hat{\xi}) - 2\ln (\Phi(\hat{\xi})) \ = \ C +2\hat{\xi}
\end{equation}
Therefore if $\Phi(\hat{\xi}) \equiv a$ \ {\rm then } \ $a - 2\ln (a) = C+2\hat{\xi}$ 
determines $a$ given $C$ and $\hat{\xi}$.


By adjusting $\mu_0$, we can always choose $\hat{\xi}$ to be zero or any
value. 

At the minimum of $\Phi$, \ $\Phi = C+2\ln (\Phi) + 2\xi $ which implies:
$$
G \ = \ \frac{\Phi}{(\Phi+1)^2} \left[ \Phi+1\right] \ = \ \frac{\Phi}{\Phi+1} < 1
$$

If $\Phi \rightarrow  0$ at any {\it finite} $\xi_0$ then $\Phi' \rightarrow 0$ 
as $\xi \rightarrow \xi_0$ and $G \rightarrow 0$ as well. Since at finite $\xi$ 
we have everything regular, the det $g \ne 0$ and therefore $\Phi$ and $G$ both can 
not be allowed to go to zero.

Thus, there can be no event horizon at any {\it finite} $\xi_0$, and 
$\Phi$ and $G$ are necessarily $> 0 ~~\forall $ finite $\xi$.

\par {\underline{Now consider the asymptotics:} } \\

Observe that $\Phi$ can not be oscillatory since it can have at the most
one extremum. Hence it is either bounded or unbounded as $\xi
\rightarrow \pm \infty$.

If $\Phi$ has a {\it {finite, non-zero}} limit, then the equation implies that
$\Phi' ~ \rightarrow ~ \pm \infty$ which is absurd. Therefore $\Phi$ {\it
{either}} vanishes {\it {or}} diverges to $\infty$.

If $\Phi ~ \rightarrow ~0$ , then we can approximate the equation for $\Phi'$
as:
$$
\Phi' ~~  \simeq 2 ~ \Phi ~ (\xi + \ln (\Phi)) ~~~~ \Rightarrow
$$
$$
\ln (\Phi) ~~ \simeq ~~ \#e^{2\xi} - \frac{1}{2} - \xi .
$$
This will go to $- \infty$ provided \# is zero and $\xi \rightarrow + \infty$.
Clearly then as $\xi \rightarrow -\infty,~ \Phi $ must diverge to infinity.

As $\xi \rightarrow +\infty , \Phi' \rightarrow 0$ which in turn implies that
$G \rightarrow 0$ as well. But $(\Phi G)' = 2 G^2$ and therefore $\Phi G$
increases monotonically and hence can {\it not} vanish. Thus $\Phi$ {it must}
diverge as $\xi \rightarrow + \infty$ as well.

To summarise, 
$\Phi \rightarrow +\infty \ {\rm as} \ \xi \rightarrow \pm \infty$ must hold
which in turn implies that $\Phi$ must have a minimum.

Consider approximate solution as $\Phi \rightarrow \infty 
(\xi \rightarrow \pm \infty)$ .

Let 
\begin{equation}
\eta \equiv \Phi-2\ln (\Phi) -2 \xi -C .
\end{equation}

Therefore
\begin{equation}
\eta' \ = \ \frac{\Phi-2}{\Phi+1} (-\eta )  -2\quad \mbox{where} \ 
\Phi \ = \ \Phi(\eta) 
\end{equation}

Expanding in powers of $1/\Phi$,
\begin{equation}
\eta ' \ = \ -\eta (1-\frac{2}{\Phi}) (1-\frac{1}{\Phi} + \frac{1}{\Phi^2} -
\frac{1}{\Phi^3} \ldots) -2 ~~~~ \forall ~ \Phi > 1.
\end{equation}

For $\Phi > > 1 , \ \eta ' \approx -\eta -2 \Rightarrow$
$$
\eta \ = \ De^{-\xi} -2 ~~~~ \Rightarrow
$$
\begin{equation}
\Phi -2\ln (\Phi) \ = \ C-2 + 2\xi + De^{-\xi}
\end{equation}
and this is consistent with $\Phi >>1$ provided either $\xi \rightarrow
+\infty$ or $\xi \rightarrow -\infty$. 

The corresponding asymptotic behaviour for $G$ can be deduced from the behaviour of
$\Phi$. The leading behaviours are given below.
$$ \begin{array}{lclcl}
\xi \rightarrow \infty & : & \Phi & \approx & 2\xi \\
& : & G & \approx & 1 + \frac{1}{2\xi} \\
& & \\
\xi \rightarrow - \infty & : & \Phi & \approx & D e^{-\xi} \\
& : & G & \approx & \frac{e^{\xi}}{D} \ (C + 1 + 2 \ln (D)) \\
\end{array}
$$

{\underline {Remark:}} The $\xi \rightarrow -\infty$ behaviour shows that $\Phi G \rightarrow
constant$. We may choose this $constant$ to be 1 by choosing $D = e^{-C/2}$.
The asymptotic form then resembles the form for a {\it negative} mass
Schwarzschild solution. This is not surprising since as $\xi \rightarrow
-\infty,~G \rightarrow 0$ and $\Phi \rightarrow \infty$ , the first terms
in the basic equations become negligible and the equations approximate
to the standard Schwarzschild equations.

To summarise :
The equations can be integrated once exactly to give
\begin{enumerate}
\item $$ G \ = \ \frac{\Phi}{(\Phi+1)^2} \left\{C+1+2\xi + 2\ln (\Phi)\right\} 
$$
$$
\Phi' \ = \ \left( \frac{\Phi}{\Phi+1}\right) \left[C+2\xi - \Phi + 2\ln (\Phi)\right]
$$
\item Regularity at finite $\xi \Rightarrow \Phi,G > 0 \ \forall$ finite
$\xi$. Therefore no event horizon is possible.
\item $\Phi$ can have at the most 1 extremum which must be a minimum.
\item $\Phi \rightarrow \infty$ as $\xi \rightarrow \pm \infty$ and therefore
$\Phi$ {\it does} have the minimum.
\item $ G' > 0$ as $\xi \rightarrow -\infty$ while $G' < 0$ as $\xi \rightarrow
\infty$ and therefore $G'$ must vanish for some finite $\xi$. $G$ then {\it
has} a unique {\it maximum}.
\end{enumerate}

This qualitative picture is corroborated by numerical integration of the $\Phi
- G$ equations as shown in the figure below. The specific initial values are
chosen for convenience only.
\begin{figure}[htb]
\centerline{
\mbox{\psfig{file=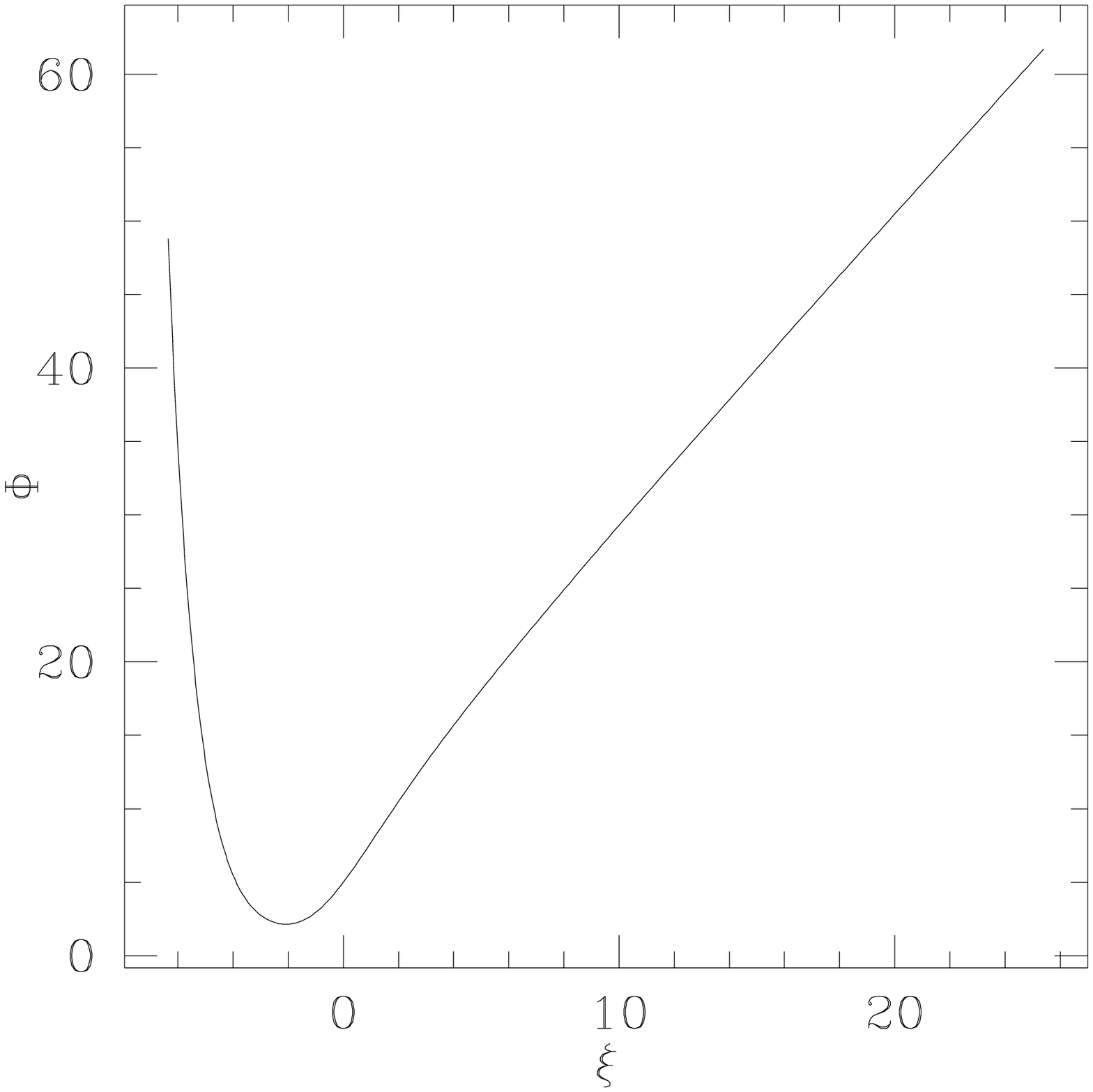,width=6.5truecm,angle=0 }}
\mbox{\psfig{file=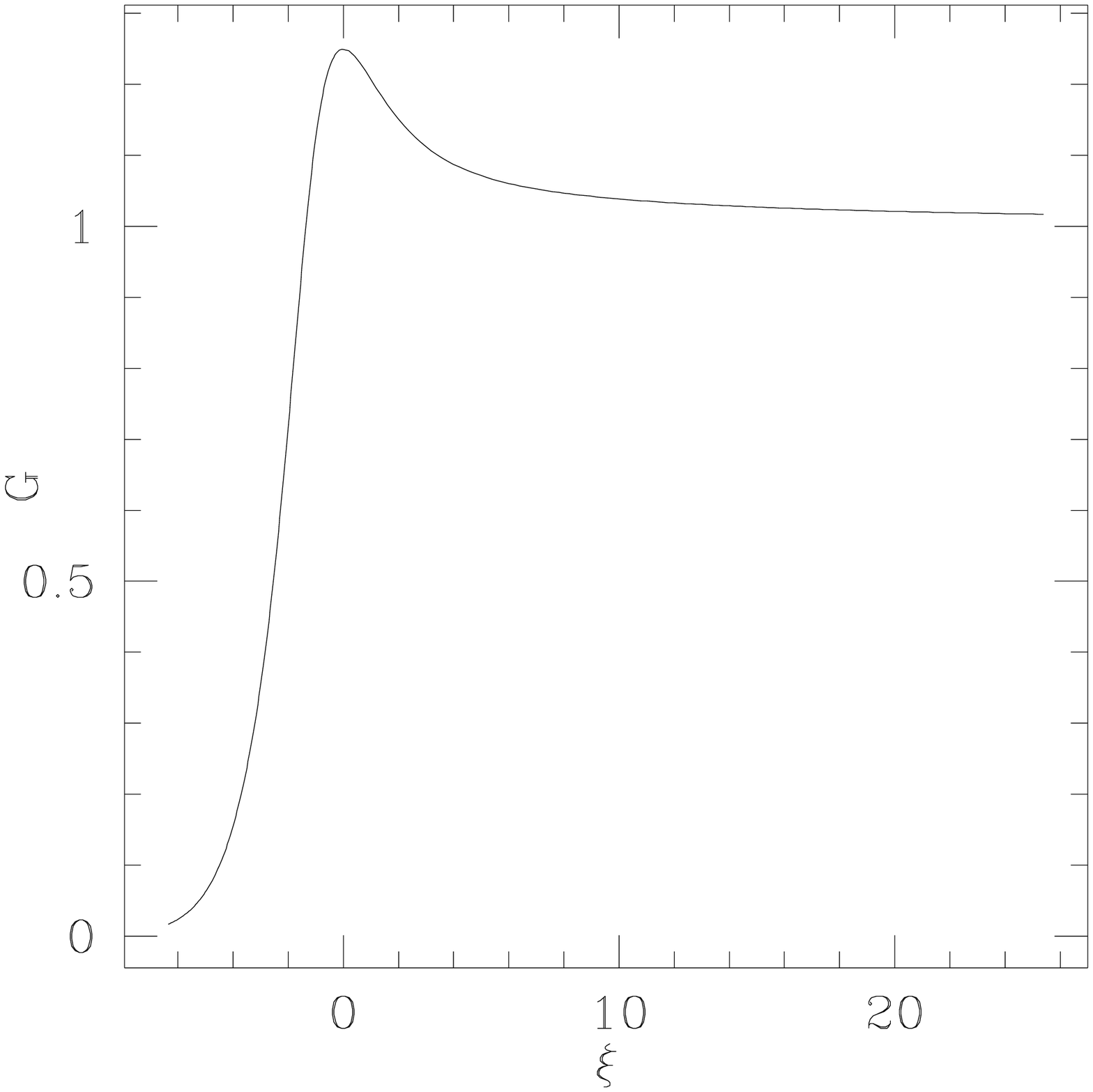,width=6.5truecm,angle=0}}
}
\caption{The solution has $G(0) = 1.25, G'(0) = 0.0$}
\end{figure}


\vskip .5cm
\centerline{ \bf{Section 3: Interpretation of the solution}}
\vskip .5cm

We have four freely specifyable parameters: the arbitrary scale $\mu_0$, the
constant of integration $\lambda$ coming from the conservation equation, the
constant $C$ and $a \equiv \Phi(0)$ (say). After the substitution $F = \lambda
\Phi$, the $\lambda$ drops out of all differential equations. 
It still appears in the {\it time} components of the vector field $k, \ell$ and the
$00$ component of the stress tensor. 
It also appears in the metric as a coefficient of the $dt^2$ term. By rescaling time 
we can remove it from the metric and from the vector fields $k, \ell$. Clearly
then its value can{\it not} have any physical meaning. In effect we take
$\lambda = 1$.

The scale $\mu_0$ on the other hand appears non trivially in the following
sense. If we take any stationary observer with four velocity 
$u^{\mu} = \xi^{\mu}/\sqrt{\Phi}$, then the energy density measured in his/her
rest frame is given by, 
\begin{equation}
\begin{array}{lcl}
\rho_u(r) & = & T_{\mu\nu}u^\mu u^\nu \\
& = &\frac{1}{8\pi r^2 \Phi(r)} \\
& = & \frac{1}{8\pi}\frac{1}{\mu_0^2} \frac{e^{-2\xi}}{\Phi(\xi)} \\
\end{array}
\end{equation}

How could a scale be chosen? As noted earlier, our solution has a curvature
singularity as $r \rightarrow 0$. So a natural approach is to consider the
solution to be valid for $r \geq R$ for some R. This $R$ then provides a
natural scale ($\mu_0 \equiv R$). We also noted earlier that the $\xi$ going to
$-\infty$ behaviour resembles that of a {\it negative} mass ($-m$) Schwarzschild
solution. The constant $D$ then equals $2m/\mu_0$. The
parameter $2m$ then provides a natural scale.  In either of the cases, a
natural scale $\mu_0$ can be chosen. It remains now to choose $C$ and $a$.

If matching with negative mass Schwarzschild solution is considered then the
choice $\mu_0 = 2m$ gives $D = 1$ or $C = 0$. The constant $a$ is left
unconstrained ($ >~ 0 $).

A more ``realistic" matching is to choose an $R$, the radius of some physical
body and match our solution with an interior Schwarzschild solution (perfect
fluid case for instance)\cite{wald}. For an
interior solution, the function $G$ is expressed in terms of a mass function
$M(r)$ 
$$ 
M(r) \equiv 4\pi \int_{0}^{r} \rho(r')(r')^2 dr' 
$$
and the equations are integrated (usually numerically). The $\Phi$
function is trivially determined once $\rho(r)$ is determined (The equation of
state gives the pressure $P(r)$ ). The matching is minimally required to have
$F, G$ and $F'$ to be continuous across the matching surfaces (See the appendix
for details). The continuity of $\Phi$ across $R$ can always be ensured trivially
since for the interior $F$ there is a constant of integration which can always
be adjusted. 

A physical body provides the following physical data namely, the physical
radius $R$ which gives $\mu_0$; the physical mass $M$ (in the absence of the
radiation shell) which gives $G(R) ~=~ (1 - 2M/R)^{-1}$; and the radiation 
density measured by a stationary observer in his/her rest frame $\rho_u(R)$ 
which gives $\Phi(R)$. As discussed in the appendix, $\rho_u(R) = P(R)$
because of continuity of $\Phi'$ .
Thus all the data needed for specifying a particular solution is available.

Thus for interior matching, we take:
\begin{equation}
{\mbox{
$\begin{array}{cclccl}
b & \equiv & G(0) & = & \frac{1}{1-\frac{2M}{R}} & \\
a & \equiv & \Phi(0) & = & \frac{1}{8 \pi R^2 \rho}
\end{array}$
}}
\end{equation}

Given $a$ and $b$, the constant $C$ is given by,

\begin{equation}
C \ = \ \frac{b (a + 1)^2}{a} - 1 - 2 \ln (a)
\end{equation}

This gives a method of choosing the constants of integration in a given
physical context, thus {\it determining} the solution appropriate for the
context.

The solution so determined is to be evolved up to some
{\it {finite}} $\bar{\xi}$ as we do not have asymptotic flatness. At this point
we would like to match our solution to an {\it {exterior}} Schwarzschild solution. 
As discussed in the appendix, it is {\it{not}} possible to do so while
maintaining the continuity of $\Phi'$ and a ``regularising" thin layer must be
added on. This can be done. Suppressing the ``thicknes" of the regularising
layer we see that the
continuity of $G$ provides us with an $\bar{M}$. However, since
$\Phi(\bar{\xi})$ is not equal to $G^{-1}(\bar{\xi})$ the exterior $F$ function
will go to $\Phi(\bar{\xi})$ as $\xi$ goes to infinity. The mass given by the
Komer integral will then have a normalization such that this mass is given by
the $\bar{M}$.

Thus for exterior matching at $\bar\xi$,  we set
\begin{equation}
G(\bar\xi) \ = \ \frac{1}{1- \frac{2\bar{M}}{R} e^{-\bar\xi}}
\end{equation}

It follows then,
\begin{equation}
\frac{\bar{M}}{M(R)} \ = \ \left[ \frac{G(\bar\xi)-1}{G(0)-1} ~~~
\frac{G(0)}{G(\bar\xi)} \right] e^{\bar\xi}
\end{equation}

Mass of such a body will be larger by factors, from the mass it would
have had in the absence at the radiation shell. (equivalently from the mass
determined from the interior dynamics).

To get a feel, let us put in some numbers. Let $\Lambda_T$ denote the energy
density of the background radiation at temperature $T$. It is given by,
$$
\Lambda_T \sim 10^{-15} \times T^4 ~~ ergs/cm^3
$$
The $\rho_u(\xi)$ on the other hand is given in conventional units by,
$$
\rho_u(\xi) = (\frac{c^4}{G_{Newton}})(\frac{1}{8\pi R^2} )(\frac{e^{-2\xi}}{\Phi})
$$ 
Therefore,
$$
e^{\xi} \sim ( \frac{c^4}{8\pi G_{Newton} \Lambda_T} )^{1/2} (\frac{1}{R \sqrt{\Phi} T^2}) 
$$
Or,
$$
e^{\xi} \sim ( 10^{31} )(\frac{1}{R \sqrt{\Phi} T^2}) 
$$
Putting $\xi = 0$ gives,
$$
a \ \sim \ ( 10^{62} )(\frac{1}{R^2 T^4})
$$
For white dwarf (say), $b$ is typically about $1 + 10^{-4}$ \cite{shapiro}. 
Typical white dwarf radius is
about $10^9$ cm. An astrophysical body such as a white dwarf could not be
expected to have been formed in the earlier epoches and thus the background
temperature can not be larger that about $10^4$. 
$a$ is then about $10^{28}$! 

Notice that in this case $a (b -1) \gg b$. As $\xi$ is increased $a$ increases
and $b$ decreases, relatively slowly, maintaining the inequality. Thus the first
terms in both of the basic equations are negligible. But then the equations
approximate the usual Schwarzschild case and $M(\xi)$ read off from $G$ will be
essentially a constant i.e. $\bar{M}/M$ will be very close to 1. Numerical
corroboration of this is shown in the figure below. 
\begin{figure}[htb]
\centerline{
\mbox{\psfig{file=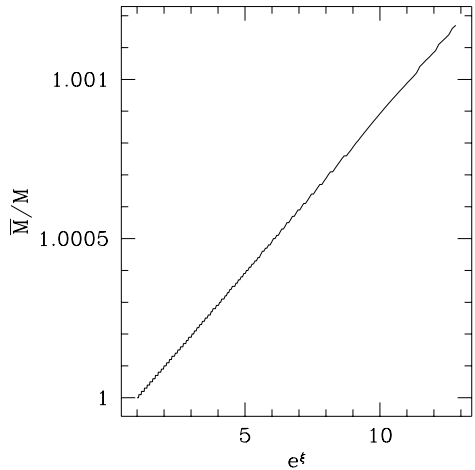,width=6.5truecm,angle=0 }}
}
\caption{The solution has $\Phi(0) = 10^{10}, G(0) = 1 + 10^{-6} $}
\end{figure}

It is also clear that to
get significant deviations from the usual Schwarzschild case one must have
$a (b - 1)/b$ to be comparable to 1 or less than 1. In any astrophysical
context (excluding black holes), the $T$ would be about the same order while 
$b -1$ continues to be not {\it too} small. Only way then to get a deviation is to 
reduce $a$ i.e. increase $R$.

Indeed if we take a spherical galaxy to be the inner body then $R \sim 10^{22}$
cms. ($10^5$ light years), the mass $M$ is about $10^{12}$ solar mass giving $b
 - 1\sim 10^{-4}$ - $10^{-5}$. $a$ is about 100 and significant deviations from the
usual Schwarzschild solution can be expected. The figure below corroborates
this expectation.
\begin{figure}[htb]
\centerline{
\mbox{\psfig{file=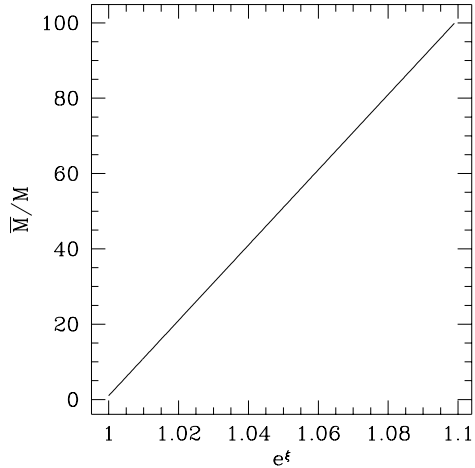,width=6.5truecm,angle=0 }}
}
\caption{The solution has $\Phi(0) = 10^{2}, G(0) = 1 + 10^{-5} $}
\end{figure}
The ratio $\bar{M}/M$ is several orders of magnitudes
bigger than 1 even for $\bar{R}/R$ slightly larger than 1 ! 
The radiation shell will of course be dark since it will be merged with the 
background radiation. The shell then seems to be a candidate for dark matter at
least in some cases.

What about the core being a black hole? Clearly we can not match our solution
at the horizon but we could try matching outside the horizon. A black hole can
provide an out going flux only via the Hawking mechanism. Since we are using
stationary observers to get the value of $a$, the energy density should be
that corresponding to the local temperature. The temperature is then given by,
$$
T \ = \ 10^{-6}\frac{M_{\odot}}{M} \sqrt{b}
$$
Setting $\alpha \equiv M/M_{\odot}$ and $\beta \equiv R/R_{Schwarzschild} $,
one gets,
$$
a(b - 1)/b \ \sim \ 10^{76} \alpha^2 \frac{(\beta -1)^2}{\beta^5}
$$
For $R$ not too large ($\beta$ of the order of 10, say) deviations are possible
only for {\it {extremely}} light black holes!

{\underline {Remark}} $~$ Quite apart from these numbers,
the precise matching of the
two rates is fragile though. As the universe expands, the incoming rate will
decrease and the black hole will begin loosing its mass, thereby {\it increasing}
its out going rate and preventing return to equilibrium. The net result of
the earlier equilibrium is perhaps to delay the evapouration process. 

The similarity of our solution to a negative mass Schwarzschild solution for
$r$ close to zero suggests a purely speculative possibility of taking the core to
be a ``negative mass" body. Of course no such body is known! If at all it
``exists" one could only imagine a quantum origin. All the scales may then be
taken to be Planckian. $a$ then is of the order of 1. Numerical solution then
indicates that the $\bar{M}$ becomes positive for $\bar{R}/R$ greater than
about 2 - 10.

Though spherical symmetry, staticity and the particular form of the stress tensor 
are obvious idealizations, one can still observe the following:
\begin{enumerate}
\item[1.] The stress tensor satisfies all the usual energy conditions and as
such is a physically possible/admissible one. The Einstein equations then lead us to a
solution discussed above. 
\item[2.] Apart from the role of providing incoming radiation, the
background radiation or its temperature appears explicitly in providing one of
the constants of integration, $\Phi(0)$. The solution then seems best interpreted
as a radiation shell near the surface of a spherical static body. This shell
however contribute to the mass significantly only for bodies with sizes on the
galactic scale. The effective darkness of such bodies indicates a possibility
for dark matter at least in some cases. 
\end{enumerate}

\vskip.6cm
{\bf {Acknowledgements:}} It is a pleasure to acknowledge useful comments on
the work by B. R. Iyer. I would also like to thank D.C. Srivatsava, Andrzej
Krasinski, N.D. Hari Dass and B.R. Iyer for help on some related literature.
Comments by one of the referees about the matching conditions used in the
section 3, lead to the inclusion of the appendix. The referee's
constructive criticism is gratefully acknowledged.

\vskip .5cm
\centerline{\bf{ Appendix }}
\vskip .5cm

\setcounter{equation}{0}
We collect here a few details of the most general equations for static,
spherically symmetric non-empty space times which are useful for discussion of
the matchings involved. 

Define the orthonormal set of vectors: 
\begin{equation}
\begin{array}{lclclcl}
e^\mu_0 & = & \frac{1}{\sqrt{F}} (1, 0, 0, 0) & ; & e^\mu_1 & = & \frac{1}{\sqrt{G}}
(0, 1, 0, 0) ; \\
e^\mu_2 & = & \frac{1}{r} (0, 0, 1, 0) & ; & e^\mu_3 & = & \frac{1}{r sin(\theta)}
(0, 0, 0, 1) . \\
\end{array}
\end{equation}

Any $T^{\mu\nu}$ can then be expressed as,
\begin{equation}
T^{\mu\nu} ~ = ~ \rho_{ab} e^\mu_a e^\nu_b , ~~~~\rho_{ab} ~ = ~ \rho_{ba}.
\end{equation}

Spherical symmetry and staticity ($R_{01} = 0$) implies that,
\begin{equation}
\rho_{ab} ~ = ~ \mbox{diag} (\rho_0(r), \rho_1(r), \rho_2(r), \rho_3(r) ),
\end{equation}

with $\rho_3(r) = \rho_2(r)$. 

The conservation equations give a single equation: 
\begin{equation}
\frac{d\rho_1}{dr} + \frac{2(\rho_1 - \rho_2)}{r} + \frac{(\rho_0 + \rho_1)}{2F}
\frac{dF}{dr} = 0 .
\end{equation}

Some special cases are: 
\begin{equation}
\begin{array}{lclcl}
\mbox{Perfect fluid} & : & \rho_0 \equiv \rho & , & \rho_1 = \rho_2 \equiv P
\\
\mbox{Reissner-Nordstrom} & : & \rho_0 = \frac{Q^2}{r^4} & , & \rho_0 = \rho_2 = -\rho_1 \\
\mbox{Present case} & : & \rho_2 = 0 & , & \rho_1 = \rho_0 \equiv \rho \\
\end{array}
\end{equation}

The Einstein equations can be organised as explained in section 1 to get a set
of three first order differential equations as: 
\begin{equation}
\begin{array}{lclcr}
\frac{dF}{dr} & = & F (G - 1) + (8\pi r^2 F\rho_1)G  & .... & (a) \\
& & & & \\
\frac{dG}{dr} & = & -G (G - 1) + (8\pi r^2 F\rho_0)\frac{G^2}{F} & .... & (b) \\
& & & & \\
\frac{d\rho_1}{dr} & = & 2 (\rho_2 - \rho_1) - \frac{\rho_0 + \rho_1}{2F}\frac{dF}{dr}
 & .... & (c) \\
\end{array}
\end{equation}
\par Defining,
$$
\sigma_a \equiv 8 \pi r^2 F \rho_a ~~~~~ a = 0, 1, 2 \\
$$

and using the dimensionless variable $\xi \equiv \ln ( r/\mu_0 )$, we get:
\begin{equation}
\begin{array}{lclcr}
\frac{dF}{d\xi} & = & F (G - 1) + \sigma_1 G  & .... & (a) \\
& & & & \\
\frac{dG}{d\xi} & = & -G (G - 1) + \sigma_0 \frac{G^2}{F} & .... & (b) \\
& & & & \\
\frac{d\sigma_1}{d\xi} & = & 2 \sigma_2 - \frac{\sigma_0 - \sigma_1}{2F}\frac{dF}{d\xi}
 & .... & (c) \\
\end{array}
\end{equation}

Mathematically we have an underdetermined system of equations with $\sigma_0$
and $\sigma_2$ (say) as freely specifiable function. Physically of course the
$\sigma_a$ 's are to be determined by the dynamics of the matter constituents
eg. Maxwell equations for the Reissner-Nordstrom case, equation of state for
the perfect fluid case and modeling in our case. 

Usually one notes that the (6b) equation does not involve $F$ and solves this
equation in terms of, 

$$
M(r) \equiv 4 \pi \int \rho_0(r') (r')^2 dr' ~ , ~ G \equiv 
(1 - \frac{2 M(r)}{r} )^{-1} \\
$$

The (6c) equation which involves only $\rho_a$ 's is then solved and (6a) equation 
can then be trivially integrated. 

Since in our case equation (6c or 7c) are trivially integrated we were able to
reduce the remaining equations to a single first order differential equation.

In fact following our way of organising the equations one can construct the
following \underline{\it{exact solution}}. 

Taking 
$\sigma_1 = \sigma_0 \equiv \sigma$ and $\sigma_2 = -A e^{-\xi}$ gives us
$\sigma = B + 2 A e^{-\xi}$ and allows the remaing two equations to be reduced
to a single {\it{first order}} equation for $F$. $A = 0$ reproduces our equation
(22) of section 2. However for $B = 0$ and $A > 0$ (to satisfy energy 
conditions) further integration gives an exact solution ($F = A \Phi , A > 0$) :
%
%
\begin{equation}
\begin{array}{lclclcl}
\Phi + 2 e^{-\xi} \ln (\Phi) & = & C + (D - 2 \xi) e^{-\xi} & , & & & \\
G & = & \frac{C}{(\Phi + 2 e^{-\xi})^2} & , &  C & > & 0 \\
\rho & = & \frac{e^{-3\xi}}{4 \pi \mu^2_0 \Phi} & , & \rho_2 & = & - \rho/2 \\
\end{array}
\end{equation}

This solution again has naked singularity and although $\rho$ falls off faster
than before, it is still not fast enough to get asymptotic flatness. 

Now we address the issue of matching our solution on the interior to a physical
body and on the exterior to an exterior Schwarzschild solution. 

It is useful to note that the energy flux, as implicit in the Komer integral,
depends on derivative of $F$. The extrinsic curvature for the hypersurfaces $t
= i$ constant and $r = $ constant also depend on the derivative of $F$ apart
from on $F, G, r$ etc. None depend on the derivative of $G$ though. If the
matching are to ensure continuity of any (or all) of these, then one must
demand continuity of $F, G$ and $F'$ across the matching spheres. The
equations 6a (or 7a) show immediately that such a matching {\it {can not}} be
done on the exterior because $\rho_1$ is not continuous across $\bar{R}$ .
On the interior though such a matching is possible. Note that the matching
requires continuity of $\rho_1$ only and NOT of $\rho_0$ .

For an interior body described by perfect fluid stress tensor we have, 

$$
\frac{dF}{d\xi}|_{\xi = 0} = 2 G(0)\frac{M(0) + 4 \pi R^3 P(0)}{R}. \\
$$

Putting $a \equiv F(0)$, $b \equiv G(0)$ $d \equiv F'(0)$ and noting that
$b = (1 - 2 M(0)/R)^{-1}$ we see that 

$$
d = a (b - 1) + 8 \pi R^2 a b P(0) ~~~~(\mbox{ for interior body} ) \\
$$

while for our solution 

$$
d = a (b - 1) + b
$$

Thus the matching implies, 

$$
a = \frac{1}{8 \pi R^2 P(0)}
$$

By comparing with equation (31) of section 3, we see that $P(0) = \rho_u(0)$.
This fixes the constant of integration for the interior solution as well as
provides an initial condition for our equations. 

In the absence of the radiation shell, $R$ is determined precisely by demanding
$P(R) = 0 $. We can not do so. So our choice must be some what smaller than the
usual value for $R$. 

Recall from section 3 that we chose $a$ by equating the energy density measured
by a stationary observer to the background energy density. In the conventional
units, $\rho_u(0)$ (and hence $P(0)$) is about $10^{-15} \times T^4 $ which is small 
enough so that 
the value of $R$ will be very close to that determined in the
absence of the radiation shell. So for estimate purposes we can use our earlier
estimates. The solution thus determined is essentially the same as in the
section 3. 

The exterior matching is to be examined now because the estimate of $\bar{M}$
depends crucially on this. 

We can modify the matching at $\bar{\xi}$ used in section 3 by adding a thin 
``regularising layer" which will match with our solution at $\bar{R}$ and match
with exterior Schwarzschild solution at slightly farther away. This matching of
course is to have continuity of $F'$ as well. 

To describe the thin layer, we consider the general equation (7). We retain
$\sigma_2 = 0$ condition but allow $\sigma_1 \neq \sigma_0$. We will choose
$\sigma_1$ suitably and solve for $\sigma_0$ using eqn(7c), i.e. 

$$
\sigma_0 = \sigma_1 - 2 \sigma_1' \frac{F}{F'} .
$$

We choose $\sigma_1$ such that: 
\begin{equation}
\begin{array}{lclclcl}
\sigma_1(\bar{\xi}) & = & 1 & ; & \sigma_1(\bar{\xi} + 2 \epsilon) & = & 0 ~;\\
\sigma_1 ' (\bar{\xi}) & = & 0 & ; & \sigma_1 ' (\bar{\xi} + 2 \epsilon) & = & 0 ~. \\
\end{array}
\end{equation}

This implies that $\sigma_0(\bar{\xi}) = 1$ and $\sigma_0(\bar{\xi} + 2 \epsilon) = 0$.
By choosing $\sigma_1$ to be monotonically decreasing we can
ensure tha energy conditions are satisfied. A simple choice is: 
\begin{equation}
\sigma_1(\xi) = \frac{1}{4} e^{-\frac{1}{2}\nu (\xi - \bar{\xi})^2} \{ (\frac{\xi - \bar{\xi} - \epsilon}{\epsilon})^3 - 3 (\frac{\xi - \bar{\xi} -
\epsilon}{\epsilon}) + 2 \} 
\end{equation}

By taking $\nu$ large one can control the decrease in $\sigma_1$ while by
taking $\epsilon$ small one can make the layer thin. At $\bar{\xi} + 2 
\epsilon$ 
one can match with the exterior Schwarzschild solution and read off mass from
the value of $G$. 

Numerical exploration of this thin layer shows that the qualitative conclusions
derived in the section 3 do not change. In particular, for stellar scales the
radiation shell contributes negligibly while for galactic scales there is
significant enhancement of the mass as detected from far away.
%
%
\vskip 1cm

\end{document}